\documentstyle[sprocl,epsfig]{article}

\def\Journal#1#2#3#4{{#1} {\bf #2}, #3 (#4)}

\def\NPB{{\em Nucl. Phys.} B}

\def\PLB{{\em Phys. Lett.}  B}

\def\PRD{{\em Phys. Rev.} D}
\def\ZPC{{\em Z. Phys.} C}

\def\PREP{{\em Phys. Rep.}}
\def\be{\begin{equation}}
\def\ee{\end{equation}}
\def\beeq{\begin{eqnarray}}
\def\eeeq{\end{eqnarray}}
\def\funp{{I\!\!P}}
\def\xp{x_{{I\!\!P}}}

\begin{document}

\title{REGGEONS IN DIFFRACTIVE INTERACTIONS IN DEEP INELASTIC
       SCATTERING AT HERA
\footnote{to be published in  Proceedings of the 
{\it Madrid low-$x$ Workshop,}
Miraflores de la Sierra, June 18-21,1997}       
}
\author{K. GOLEC-BIERNAT$^{a,b}$, J.KWIECI\'NSKI$^{a}$ {\it and} A.SZCZUREK$^{b}$
\medskip
%\footnote{on leave of absence from 
%Institute of Nuclear Physics,
%Krak\'ow, Poland}
}

\address{$^{a}$Department of Physics,
University of Durham, Durham DH1 3LE, England\\
\smallskip
$^{b}$Institute of Nuclear Physics, Radzikowskiego 152, 
Krak\'ow 30-342, Poland}

%\author{ A.N. OTHER }
%\address{$^{b}$Institute of Nuclear Physics, Radzikowskiego 152, 
%Krak\'ow 30-342, Poland}

%%%%%%%%%%%%%%%%%%%%%%%%%%%%%%%%%%%%%%%%%%%%%%%%%%%%%%%%%%%%%%
% You may repeat \author \address as often as necessary      %
%%%%%%%%%%%%%%%%%%%%%%%%%%%%%%%%%%%%%%%%%%%%%%%%%%%%%%%%%%%%%%

\maketitle\abstracts{
The "triple-Regge" analysis of 
subleading Reggeon contributions to 
the diffractive structure function in DIS at HERA is presented. 
The recently published data allow
to determine  the only free parameter of the analysis related to
the ratio of the  $RR\funp$ and  $\funp\funp\funp$ couplings. 
The large value of this ratio is prefered which is in agreement with 
analyses of soft hadronic interactions.
The role of the subleading Reggeons as well as pions
for the fast forward neutron production in diffractive 
processes is estimated. The forward $\pi N$ state production is
also discussed.
}

%%%%%%%%%%%%%%%%%%%%%%%%%%%%%%%%%%%%%%%%%%%%%%%%%%%%%%%%%%%%%%
\section{Introduction}

The new diffractive DIS data from  HERA, 
published by the H1 collaboration\cite{newdat}, 
indicate breaking of the Regge factorization of the diffractive
structure function. The physical picture which lies behind 
this factorization is the "soft" Pomeron emission from the proton
and a subsequent hard scattering of the virtual photon on a parton
in the Pomeron \cite{olddat,CKMT95,GK95}, (for the alternative
description see \cite{NZ92,BH95}).

In order to describe the breaking of the factorization
it was recently proposed
to include the contribution coming  from subleading Reggeons
\cite{newdat,GK96,NIK2}. In the 
approach presented here the diffractive structure function is written 
as \cite{GK96}:
\begin{equation}
\frac{dF_2^{D}}{dx_{\funp} dt}(x,Q^2,x_{\funp},t) =
f^{\funp}(x_{\funp},t) \, F_2^{\funp}(\beta,Q^2) \, + \,
\sum_{R} f^{R}(x_{\funp},t) \, F_2^R(\beta,Q^2) \; ,
\label{diff4_PR}
\end{equation}
where $\beta=x/x_{\funp}$, and $f^{\funp}$ and $f^{R}$ are the Pomeron and
Reggeon flux factors respectively, and 
$F_2^{\funp}$ and $F_2^R$ are their DIS structure functions
\footnote{The term "diffractive processes" applies
only to processes described by the Pomeron exchange.
For simplicity we shall use the same terminology for the
Reggeon exchanges.}
%including processes with the forward neutron in
%the final state which correspond to $I=1$ exchange.}. 
The details of the
Pomeron contribution can be found in \cite{GK95}, and we shall concentrate
in this presentation on the Reggeon part of (\ref{diff4_PR}).

The Reggeon flux factors are parametrized in analogy to the Pomeron case
\begin{equation}
f^R(x_{\funp},t) =
%C(s_{\gamma p} = W^2) \;
  {N \over 16 \pi} \;
 x_{\funp}^{1-2\alpha_R(t)} \;
 B_{R}^2(t) \; |\eta_R(t)|^2 \;,
\label{reg_flux}
\end{equation}
where  $\alpha_R(t)$
%=\alpha_R(0)+\alpha_R^{'}~t$ 
is the Reggeon trajectory,
$B_{R}(t) = B_{R}(0) \exp({t/{(2 \Lambda_R^2)}})$ with
$\Lambda_R=0.65~GeV$,
as known from the Reggeon phenomenology of hadronic reactions\cite{Regge}, 
describes the coupling of the Reggeon to the proton.
The function $\eta_R(t)$ is  the  signature factor\cite{Collins}: 
$|\eta_R(t)|^2 =4\,\cos^2(\pi \alpha_R(t)/2)$ or
$4\,\sin^2(\pi \alpha_R(t)/2)$
for even and odd signature Reggeons, respectively.

The exchange (emission) of  the isoscalar Reggeons
$(f_2,\omega)$ dominates the Reggeon contribution to 
(\ref{diff4_PR}), and the isovector subleading Reggeons $(a_2,\rho)$
can  be neglected \cite{GK96}. However, they  become
important for the processes with leading neutron in the final state.
Therefore we include them in our analysis and show 
that for small values of $x_{\funp}$ the isovector Reggeons give the
dominant contribution to  the leading neutron production.
 
The size of the Reggeon contributions is determined by the
Reggeon-proton couplings $B_R(0)$. They can be estimated from 
the energy dependence of the total hadronic cross section, see \cite{GKS97}
for the description of the method. As a result we obtain that
$B_{f_2}^2(0) =75.49~mb$, $B_{\omega}^2(0) =20.06~mb$, 
$B_{a_2}^2(0) =1.75~mb$ and $B_{\rho}^2(0) =1.09~mb$, where 
the same Reggeon trajectory, $\alpha_R(t)=0.55+1.0~t$, was assumed
for the four Reggeons. One clearly sees the following ordering
\begin{equation}
B_{f_2}^2(0) > B_{\omega}^2(0) \gg B_{a_2}^2(0) \sim B_{\rho}^2(0) \; ,
\label{order}
\end{equation}
which confirms the dominance of the isoscalar Reggeons. 
%over the 
%isovector ones. 

We are interested in the  Reggeon structure function $F_2^R$
for small $\beta$, since we expect that in this   kinematical
limit the Reggeon contribution to (\ref{diff4_PR})
is most important. In that limit we can apply the 
"triple-Regge" analysis~\cite{FFOX,Regge} to our problem,  
(see \cite{GK95,GK96} form more details), in which 
both the Pomeron and Reggeon structure functions are of the
form
\begin{equation}
F_2^{\funp,R}(\beta) = A_{\funp,R} \; \beta^{-0.08}  \; ,
\label{pbeta}
\end{equation}
for small $\beta$, where the ratio \begin{equation}
\label{cenh}
C_{enh} = \frac{A_R}{A_\funp}~
\end{equation}
is related to the ratio of  the "triple-Regge" $R R \funp$ and
$\funp \funp \funp$ couplings, 
and should be much bigger than one, as suggested by the analysis \cite{FFOX}
of soft hadronic interactions. 

Indeed the new data from HERA \cite{newdat}, presented in terms
of the structure function 
(\ref{diff4_PR}) integrated over $t$ and denoted by $F_2^{D(3)}$,
prefer $C_{enh}\approx10$ 
in which case a reasonable agreement of our description with the data
is obtained for $\beta \leq 0.4$. 
This is illustrated in Fig.1  where the pure Pomeron contribution 
from\cite{GK95} (solid lines)
and the effect of the Reggeon terms (dashed lines) is shown.

\begin{figure}[p]
   \vspace*{-1cm}
    \centerline{
     \psfig{figure=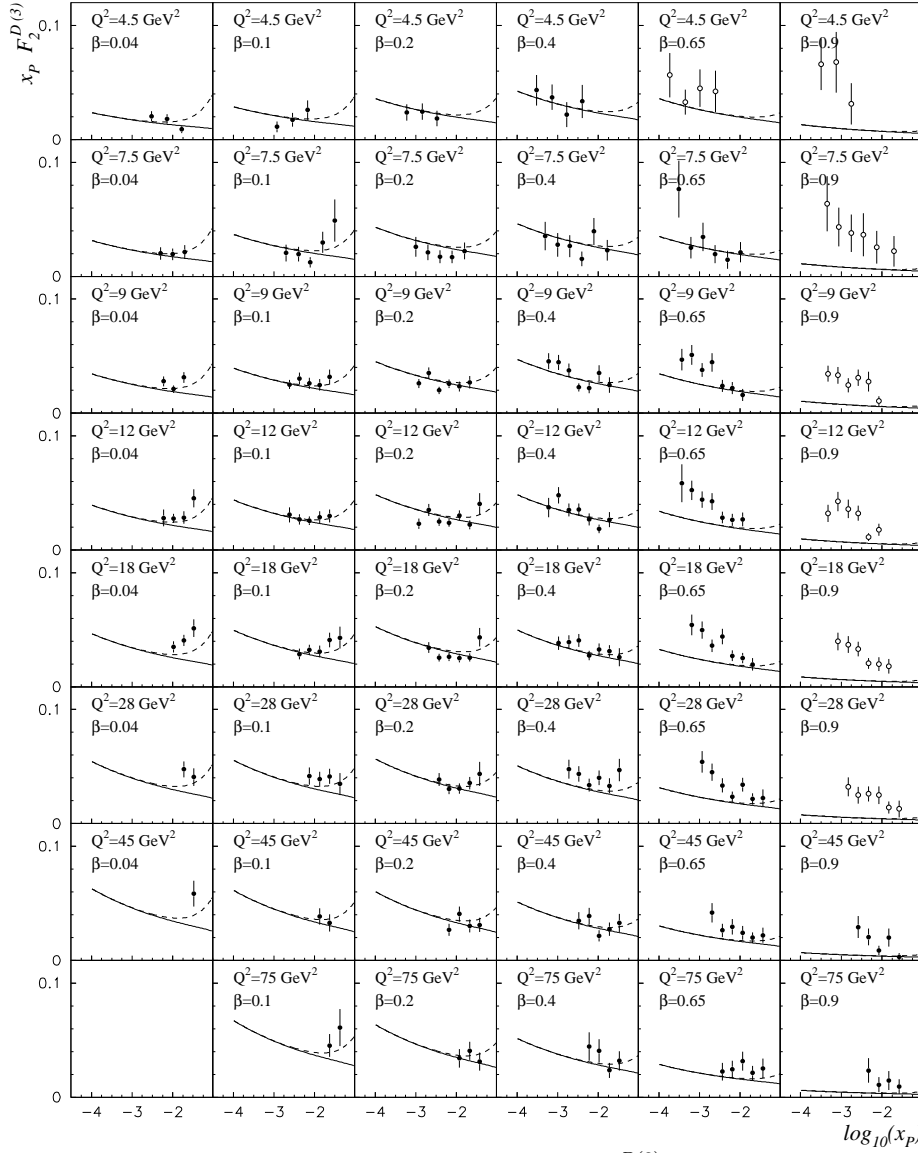,height=16.25cm,width=13cm}
               }
    \vspace*{-0.5cm}
     \caption{ 
The comparison of the
structure function $x_{\funp} F_2^{D(3)}$  measured at HERA and the results of
the presented analysis. The solid lines correspond to the pure 
Pomeron contribution  while
the dashed lines show the effect of the Reggeon exchanges 
for $C_{enh}=10$.
}
\end{figure}

%%%%%%%%%%%%%%%%%%%%%%%%%%%%%%%%%%%%%%%%%%%%%%%%%%%%%%%%%%%%%%%%%%%%%
\section{Fast forward neutron production}

\begin{figure}
   \vspace*{-1cm}
    \centerline{
     \psfig{figure=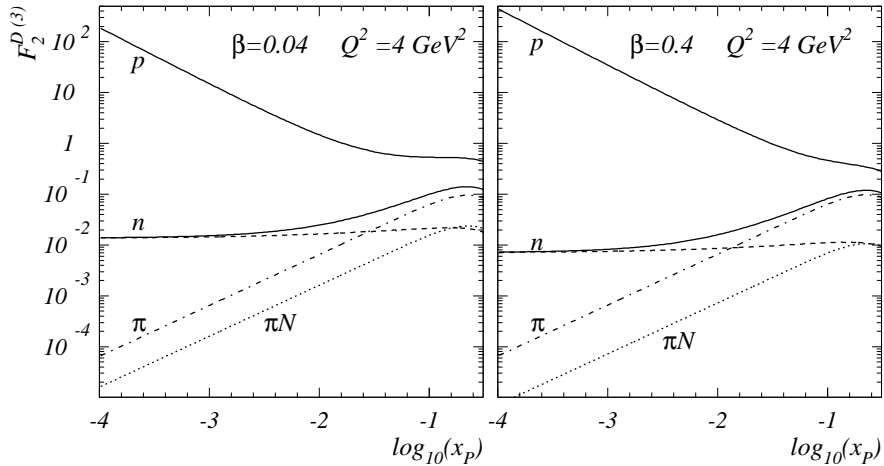,height=7.8cm,width=13cm}
               }
    \vspace*{-0.5cm}
     \caption{ 
The proton and neutron contributions 
to $F_2^{D(3)}$ for $Q^2 =4 GeV^2$ and $\beta=0.04$ and $0.4$.
The $(a_2,\rho)$ Reggeon (dashed lines)
and $\pi^{+}$ (dot-dashed lines) contributions 
to $\Delta^{(n)}$ are also shown. The $\pi N$ contribution 
to  $\Delta^{(n)}$ is shown
as the dotted line.
}
\end{figure}

The  model presented here allows to separate the diffractive
structure function $F_2^{D(3)}$ into two distinct contributions
$\Delta^{(p)}$ and $\Delta^{(n)}$ with the leading proton or neutron
observed in the final state, respectively
\begin{equation}
F_2^{D(3)}(\beta,x_{\funp},Q^2) =
\Delta^{(p)}
%F_2^{D(3)}
(\beta,x_{\funp},Q^2) +
\Delta^{(n)}
%F_2^{D(3)}
(\beta,x_{\funp},Q^2) \; .
\label{pncontri}
\end{equation}
Additionally, we add pions to our model. Their contribution to (\ref{diff4_PR})
has the same form as the Reggeon contribution, but now the one-pion exchange
model was assumed for the form of the pion flux factor 
$f_{\pi}(x_{\funp},t)$ and
the GRV parametrization\cite{GRV} of the pion structure function
$F_2^{\pi}(\beta, Q^2)$ was used \cite{GKS97,PSI96}.

In Fig.2 we show $\Delta^{(p)}$ and $\Delta^{(n)}$ marked by the solid lines.
As expected the proton contibution dominates
over the neutron one almost in the whole range of $x_{\funp}$ because
of the Pomeron contribution present in $\Delta^{(p)}$ but absent
in $\Delta^{(n)}$.
The ratio $\Delta^{(n)}/\Delta^{(p)}$ becomes significant $(\sim 0.1)$
only for $x_{\funp} > 0.1$, where the Pomeron exchange
is suppressed. In this case the Reggeon and pion contributions
come into play. However,  this region needs a
careful treatment since it might already be unsuitable for
the Regge analysis.

The lower curves in Fig.2 show different contributions to the
fast forward neutron production. For large values of
$x_{\funp}$ the $\pi^{+}$  exchange process,
marked by the dash-dotted line,
is the dominant effect.
The situation changes when $x_{\funp}$ is getting
smaller and  the Reggeon exchanges (shown by  the dashed lines)
are almost entirely responsible for
the fast neutron production.

Is it possible to identify these contributions experimentally?
We  suggest studies with the help of the Monte Carlo models which
are supposed to describe diffractive interaction in DIS at HERA
with the Reggeon contribution included.

%%%%%%%%%%%%%%%%%%%%%%%%%%%%%%%%%%%%%%%%%%%%%%%%%%%%%%%%%%%
\section{Diffractive $\pi N$ production}

In the presented analysis we have neglected the contribution of diffractively
produced $\pi N$ and $\pi \pi N$ states. For small values of $x_{\funp}$
relevant here only the $\pi N$ contribution is of interest \cite{HNSSZ96}.

The  analysis of the  $\pi N$ contribution to the structure function
(\ref{diff4_PR}) 
is performed in \cite{GKS97} using 
the Deck mechanism \cite{AG81}. The dominant process is the emission
of the virtual pion from the proton which couples  to the Pomeron.
In the simplest approximation the corresponding contribution can be written 
as a product of the pion flux factor $f_{\pi}$ 
and the effective structure function
\begin{equation}
F_{2}^{{\funp}/\pi}(\beta,Q^2) =
\int_{\beta}^{1} dz \int_{-\infty}^{t'_{max}} dt'
\; f_{\funp/\pi}(z,t') F_{2}^{\funp}(\tilde{\beta},Q^2) \; ,
\end{equation}
where $\tilde{\beta} \equiv \beta/z$ and $f_{\funp/\pi}=2/3~f_{\funp}$
is the Pomeron flux in the pion.
The $\pi N$ contribution to $\Delta^{(n)}$
is shown in Fig.2 as the dotted line and it 
can safely be neglected for
the forward nucleon production, especially for large $\beta$.

An interesting feature of the Deck mechanism is that it contributes
to the rapidity gap events for large $x_{\funp}$, while the pion exchange
(Sulivan process) is not expected to give rapidity gaps. Since in the
description of both processes the same pion flux factor is present,
the Deck mechanism can explain an approximately constant ratio in  $x_{\funp}$
of the number of events with  
and without  the rapidity gap and the forward nucleon,
observed by the ZEUS Collaboration\cite{Cart}.

%%%%%%%%%%%%%%%%%%%%%%%%%%%%%%%%%%%%%%%%%%%%%%%%%%%%%%%%%%%
\section{Conclusions}

The isoscalar Reggeon $(f_2,\omega)$ exchanges can describe 
the Regge factorization breaking in
the inclusive DIS diffractive processes observed at HERA.
The size of this contribution, 
determined by the HERA data, is in  agreement with 
soft hadronic reaction analyses.
The isovector Reggeons $(a_2,\rho)$ dominate the fast forward 
neutron production at small $\xp$, while for
large  $\xp$ the ${\pi}^{+}$ exchange is equally important or even dominates. 
The $\pi N$ production estimated from the Deck mechanism
is small for the forward nucleon production but 
it can explain an approximately constant ratio of the number
of  the fast nucleon events with and without the rapidity gap.

\section*{Acknowledgments}
K.G-B would like to thank the organizers of the Madrid meeting for 
the creation of 
such an enjoyable atmosphere and Beno List for  an interesting discussion. 
This work has been supported in part by the
KBN grants
no.2 P03B 231 08 and 2 P03B 184 10, 
EU under contracts Nos. CHRX-CT92-0004 
and CHRX-CT93-357, and by the grants from the
Royal Society and PPARC.

%%%%%%%%%%%%%%%%%%%%%%%%%%%%%%%%%%%%%%%%%%%%%%%%%%%%%%%%%%%%%%%%%%%%%%%

\end{document}